\documentclass[pdflatex,sn-nature]{sn-jnl}

\usepackage{graphicx}
\graphicspath{{./}{figures/}}
\usepackage{multirow}
\usepackage{amsmath,amssymb,amsfonts}
\usepackage{amsthm}
\usepackage{mathrsfs}
\usepackage[title]{appendix}
\usepackage[dvipsnames]{xcolor}
\usepackage{textcomp}
\usepackage{manyfoot}
\usepackage{booktabs}
\usepackage{algorithm}
\usepackage{algorithmicx}
\usepackage{algpseudocode}
\usepackage{listings}
\usepackage[acronym]{glossaries-extra}
\usepackage{bm}
\usepackage{xspace}
\usepackage{booktabs}
\usepackage{tabularx}
\newcolumntype{C}{>{\centering\arraybackslash}X}
\usepackage[mathlines]{lineno}

\setabbreviationstyle[acronym]{long-short}
\newacronym{gw}{GW}{gravitational wave}
\newacronym{ligo}{LIGO}{Laser Interferometer Gravitational Wave Observatory}
\newacronym{lvk}{LVK}{LIGO--Virgo--KAGRA}
\newacronym{kagra}{KAGRA}{Kamioka Gravitational Wave Detector}
\newacronym{cbc}{CBC}{compact binary coalescence}
\newacronym{em}{EM}{electromagnetic}
\newacronym{bbh}{BBH}{binary black hole}
\newacronym{imbh}{IMBH}{intermediate-mass black hole}
\newacronym{fap}{FAP}{false-alarm probability}
\newacronym{snr}{SNR}{signal-to-noise ratio}
\newacronym{gr}{GR}{general relativity}

\let\keeptoday\today

\IfFileExists{gitID.txt}{
	
	\def\today{\keeptoday, GitID {\normalfont\input{gitID.txt}}}
}{
	\def\today{\keeptoday}
}

\glsdisablehyper

\begin{document}

\title{
	On the evidence of lensing in GW231123
}

\author*[1,2,3]{\fnm{Anna} \sur{Liu}}\email{aliu3@perimeterinstitute.ca}
\equalcont{These authors contributed equally to this work.}
\author*[4,5]{\fnm{Mick} \sur{Wright}}\email{m.j.wright@uu.nl}
\equalcont{These authors contributed equally to this work.}
\author*[6]{\fnm{Jef} \sur{Heynen}}\email{jef.heynen@uclouvain.be}
\author[9]{\fnm{Eungwang} \sur{Seo}}\email{e.seo.1@research.gla.ac.uk}
\author[7,8]{\fnm{Isaac C. F.} \sur{Wong}}\email{chunfung.wong@kuleuven.be}
\author[6]{\fnm{Disha} \sur{Hegde}}\email{disha.hegde@uclouvain.be}
\author[3]{\fnm{Laura E.} \sur{Uronen}}\email{laura.uronen@link.cuhk.edu.hk}
\author[4,5]{\fnm{Harsh} \sur{Narola}}\email{h.b.narola@uu.nl}
\author[4,5]{\fnm{Chris} \sur{Van Den Broeck}}\email{vdbroeck@nikhef.nl}
\author[9]{\fnm{Martin} \sur{Hendry}}\email{Martin.Hendry@glasgow.ac.uk}
\author[7,8]{\fnm{Tjonnie G. F.} \sur{Li}}\email{tjonnie.li@kuleuven.be}
\author[3]{\fnm{Otto A.} \sur{Hannuksela}}\email{oahannuksela@cuhk.edu.hk}
\author[6,10]{\fnm{Justin} \sur{Janquart}}\email{justin.janquart@uclouvain.be}

\affil[1]{\orgname{Perimeter Institute for Theoretical Physics}, \orgaddress{\street{31 Caroline St N}, \city{Waterloo}, \state{ON} \postcode{N2L 2Y5}, \country{Canada}}}
\affil[2]{\orgdiv{Illinois Center for Advanced Studies of the Universe, Department of Physics}, \orgname{University of Illinois at Urbana-Champaign}, \orgaddress{\street{Urbana}, \postcode{61801}         \city{Illinois}, \country{United States}}}
\affil[3]{\orgdiv{Department of Physics}, \orgname{The Chinese University of Hong Kong}, \orgaddress{\street{Shatin}, \postcode{NT} \city{Hong Kong}}}
\affil[4]{\orgdiv{Institute for Gravitational and Subatomic Physics (GRASP), Department of Physics}, \orgname{Utrecht University}, \orgaddress{\street{Princetonplein 1}, \postcode{3584 CC}, \city{Utrecht}, \country{The Netherlands}}}
\affil[5]{\orgdiv{Nikhef---National Institute for Subatomic Physics}, \orgaddress{\street{Science Park}, \postcode{1098 XG}, \city{Amsterdam}, \country{The Netherlands}}}
\affil[6]{\orgdiv{Universit\'{e} Catholique de Louvain, Centre for Cosmology, Particle Physics and Phenomenology, CP3}, \orgaddress{\street{Chemin du Cyclotron 2}, \postcode{1348}, \city{Louvain-la-Neuve}, \country{Belgium}}}
\affil[7]{\orgname{KU Leuven}, \orgdiv{Department of Electrical Engineering (ESAT), STADIUS Center for Dynamical Systems, Signal Processing and Data Analytics}, \orgaddress{\street{Kasteelpark Arenberg 10}, \postcode{3001} \city{Leuven}, \country{Belgium}}}
\affil[8]{\orgdiv{Leuven Gravity Institute}, \orgname{KU Leuven}, \orgaddress{\street{Celestijnenlaan 200D box 2415}, \postcode{3001} \city{Leuven}, \country{Belgium}}}
\affil[9]{\orgdiv{SUPA, School of Physics and Astronomy}, \orgname{University of Glasgow}, \orgaddress{\street{University Avenue}, \postcode{G12 8QQ} \city{Glasgow}, \country{United Kingdom}}}
\affil[10]{\orgname{Royal Observatory of Belgium}, \orgaddress{\street{Avenue Circulaire, 3}, \postcode{1180} \city{Uccle}, \country{Belgium}}}

\abstract{
	Just like light, gravitational waves may be gravitationally lensed by massive
	objects along their travel path, however, this has not been observed so far.
	GW231123 is the most massive binary black-hole merger observed to-date, and
	quasi-circular gravitational wave analyses infer high masses and spin
	parameters that are particularly sensitive to the waveform model used. Here
	we investigate whether gravitational lensing can account for these features.
	We model the signal as a superposition of multiple overlapping images of the
	same merger and find support for lensing across four waveform models. The
	lensing interpretation reduces the support for these high spins, bringing
	both the inferred spins and masses into closer agreement across waveform
	models. A comparison with a background of unlensed signals gives the lensing
	interpretation an estimated significance of $3.96^{+0.68}_{-0.30}\sigma$.
	This makes lensing a compelling interpretation of this event but is
	insufficient for a definitive claim.
}


\maketitle

GW231123\_135430~\citep{LIGOScientific:2025rsn}---hereafter GW231123---was
detected by the two LIGO observatories~\citep{LIGOScientific:2014pky} operated
by the \gls{lvk}~\citep{LIGOScientific:2014pky, VIRGO:2014yos,
	akutsu2021overview} Collaboration on the $23^{\textrm{rd}}$ November 2023.
Under quasi-circular analyses, it is interpreted as the most massive \gls{bbh}
merger observed to date with very high component spins placing it at the tails
of the population distribution~\citep{LIGOScientific:2025pvj}.  However, this
interpretation is complicated by the short duration of the signal and by
substantial inconsistencies between waveform models. In particular, the
inferred source parameters depend strongly on the choice of waveform model.
This dependence has been attributed to the event lying in a region of parameter
space where waveform models are less well calibrated against
numerical-relativity simulations~\citep{Bini:2026kwz}. Another proposed cause
of this dependence is that it reflects limitations of the assumptions
underlying the waveform models. This could be caused by unmodelled data-quality
issues~\citep{Ray:2025rtt} or missing physics~\citep{Cuceu:2025fzi,
	Hu:2025lhv}.

One possibility in the latter category is that GW231123 was gravitationally
lensed---deflected and magnified by an intervening mass along the line of
sight. Gravitational lensing is well established for electromagnetic radiation
but has not yet been confirmed for \glspl{gw}~\citep{LIGOScientific:2023bwz, LIGOScientific:2025cwb}.
Initial interest in this hypothesis for this event was reported by the \gls{lvk}
Collaboration~\citep{LIGOScientific:2025rsn, LIGOScientific:2025cwb}, who found
stronger evidence for an isolated point-mass lens in GW231123 than for any other event
detected as of GWTC-4.0~\citep{LIGOScientific:2025slb}. Subsequent independent
analyses have found varying levels of support for the lensing hypothesis with
waveform model uncertainty remaining a prevalent issue~\citep{Goyal:2025eqo,
	Shan:2025dcd, Chakraborty:2025pxt, Chan:2025kyu}. In this work, we consider a
model-agnostic class of image configurations which enables a broad search for lensing signatures
beyond specific lens models.
Furthermore, similarly to the \gls{lvk}
analysis~\citep{LIGOScientific:2025cwb}, we consider real detector noise rather
than Gaussian noise to allow consideration of the impact of transient noise
effects within the detector.

If confirmed as a lensed \gls{gw} event, GW231123 would open significant
avenues for further scientific research. Gravitational lensing of \glspl{gw}
offers a fundamentally new probe of matter on small scales, in previously
unconstrained regions of parameter space.  In particular, future observations
of lensed \glspl{gw} could provide a new way to study compact lenses and
small-scale structure, including lens populations that are difficult to access
with electromagnetic observations. Whilst claims of gamma-ray burst lensing by
lower-mass objects exist~\citep{Paynter_2021, levan2025gravitational},
uncertainties in burst physics prevent firm conclusions. In contrast, the
millisecond temporal resolution of \gls{gw} detectors enables sensitivity to
lensing-induced time delays from lenses as light as $10^{2}M_{\odot}$. A robust
identification of lensing in GW231123 would therefore constitute the first
foray into this important probe of dark matter and structure formation (see
e.g. Ref~\citep{Goyal:2025eqo, vujeva2025dark}).

In this work, we present the first comprehensive model-agnostic analysis of the
lensing hypothesis for GW231123, finding the strongest support for this
hypothesis observed so far. We adopt the phenomenological framework introduced
in Liu et al.~\citep{Liu:2023ikc}, in which lensing is described as a
superposition of multiple \gls{gw} images with independent amplitudes and time
delays. This approach captures a broad class of physically plausible lensing
configurations such as primordial black holes, intermediate-mass black holes in
globular clusters, stellar fields, or dark matter
subhaloes without committing to a particular mass distribution. Its flexibility
allows us to systematically explore the full space of lensing signatures
consistent with the data, thereby providing the most general test to date of
the lensing interpretation of GW231123. We find an estimated significance of
$3.96^{+0.68}_{-0.30}\sigma$ for the lensing hypothesis, this is compatible
with those significance estimates from other
analyses~\citep[e.g.][]{Chan:2025kyu}  and whilst strong is insufficient for a
claim of detection.

\section*{Results}\label{sec:results}
\subsection*{Evidence for lensing signatures in GW231123}

\begin{figure}[h]
	\centering
	\includegraphics[width=\textwidth]{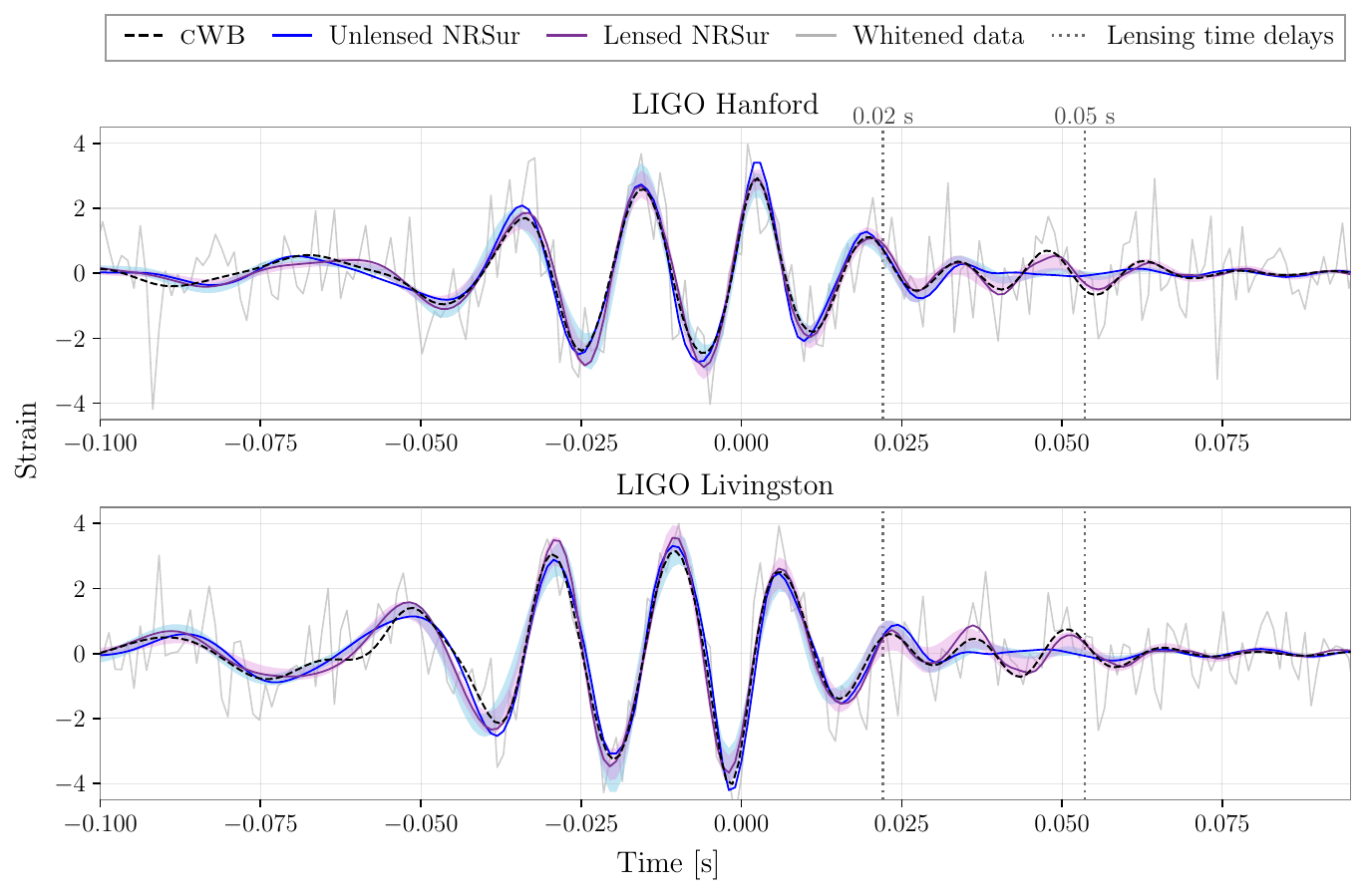}
	\caption{
		Time-domain reconstruction of GW231123 in the two LIGO
		detectors. The grey curves show the whitened strain data, while the black
		dashed curves denote the coherent power reconstruction obtained with the
		waveform model-agnostic \textsc{cWB} algorithm~\citep{Drago:2020kic}. The blue and purple curves show the
		maximum likelihood reconstructed waveforms inferred from Bayesian parameter
		estimation with the \textsc{NRSur} waveform model under the unlensed and lensed
		hypotheses, respectively; shaded regions indicate the corresponding 90\%
		credible intervals. The lensed reconstruction exhibits a better qualitative
		match to the post-merger signal morphology from $\sim 0.02\mathrm{s}$
		onwards. Vertical dotted lines mark the approximate lensing time delays
		inferred from the lensed analysis. Times are shown relative to the detection time of the signal, corresponding to 2023 November 23, 13:54:30.619 UTC.
	}
	\label{fig:waveform}
\end{figure}

Analysis of GW231123 reveals support for gravitational lensing under
the phenomenological model. To assess the robustness of
this interpretation, we analyse the signal using four waveform models:
\textsc{NRSur7dq4} (\textsc{NRSur})~\citep{Varma:2019csw},
\textsc{IMRPhenomXO4a} (\textsc{XO4a})~\citep{Thompson:2023ase},
\textsc{IMRPhenomXPHM-SpinTaylor} (\textsc{XPHM-ST})~\citep{Pratten:2020ceb,
	Colleoni:2024knd}, and \textsc{IMRPhenomXPNR}
(\textsc{XPNR})~\citep{Hamilton:2025xru}. Each of these models incorporates
higher-order multipoles and spin-precession effects. For each waveform model,
we compare the Bayesian evidence between the lensed and unlensed hypotheses and
assess the values of the lensing parameters. A summary is given in
Table~\ref{tab:bayes_factors}. We also show the reconstructed waveform under
the lensed and unlensed hypotheses for
\textsc{NRSur} in Fig.~\ref{fig:waveform}.

\begin{table*}
	\centering
	\caption{
		$\log_{10}$ Bayes factors comparing the lensed and unlensed hypotheses for
		GW231123 for different waveform models. All models favour the lensed
		hypothesis ($\log_{10}\mathcal{B}_{\mathrm{U}}^{\mathrm{L}} > 0$). For each
		model, we also report the favoured number of overlapping signals ($\# \,
			\mathrm{Images}$) and the recovered relative magnifications ($\mu_{ji}$)
		and time delays ($\Delta t_{ji}$), where the subscripts denote the image
		$j$ and its reference image $i$. Magnifications and time delays are quoted
		as posterior medians with $90\%$ credible intervals, computed from the
		posterior samples containing at least two signals ($\mu_{21}$, $\Delta
			t_{21}$) or at least three signals ($\mu_{31}$, $\Delta t_{32}$).
	}
	\label{tab:bayes_factors}
	\small
	\renewcommand{\arraystretch}{1.35}
	\begin{tabular}{lcccccc}
		\hline
		Waveform
		                 & $\log_{10} \mathcal{B}_{\mathrm{U}}^{\mathrm{L}}$ & $\# \, \mathrm{Images}$ & $\mu_{21}$             & $\mu_{31}$             & $\Delta t_{21}$ [ms] & $\Delta t_{32}$ [ms] \\
		\hline
		\textsc{XO4a}    & 0.7                                               & 2                       & $0.24^{+0.23}_{-0.22}$ & $0.04^{+0.07}_{-0.03}$ & $22^{+31}_{-7}$      & $38^{+7}_{-24}$      \\
		\textsc{XPNR}    & 2.84                                              & 3                       & $0.29^{+0.23}_{-0.17}$ & $0.04^{+0.06}_{-0.03}$ & $21^{+7}_{-7}$       & $40^{+6}_{-11}$      \\
		\textsc{NRSur}   & 3.05                                              & 3                       & $0.27^{+0.22}_{-0.15}$ & $0.04^{+0.05}_{-0.03}$ & $22^{+6}_{-7}$       & $38^{+7}_{-10}$      \\
		\textsc{XPHM-ST} & 4.65                                              & 3                       & $0.31^{+0.21}_{-0.15}$ & $0.04^{+0.06}_{-0.03}$ & $21^{+5}_{-7}$       & $39^{+7}_{-12}$      \\
		\hline
	\end{tabular}
\end{table*}

\begin{figure}
	\centering
	\includegraphics[width=\textwidth]{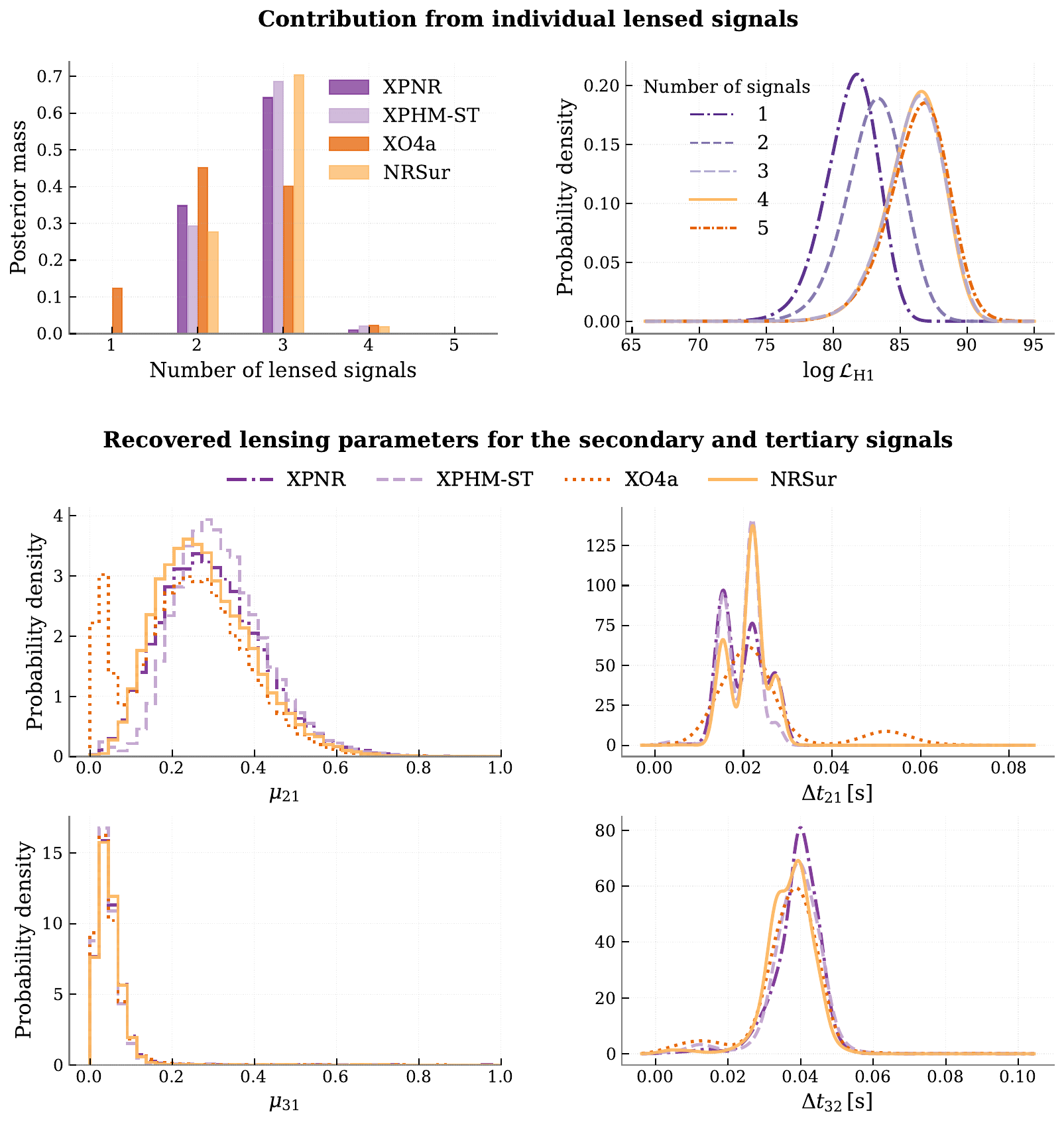}
	\caption{\textbf{Extended data}
		\textit{Top:} Posterior distributions for the recovered number of lensed signals
		inferred using multiple waveform models (left), together with the Hanford likelihood
		obtained from analyses with a fixed number of signals using \textsc{XPNR} (right).
		All waveform models favour configurations
		containing three lensed signals, with the exception of \textsc{XO4a}, which
		shows slightly higher support for two signals.
		The fixed-signal likelihoods improve from one to three signals, beyond which they
		converge. Results for the Livingston detector are qualitatively similar.
		\textit{Bottom:} Posterior
		distributions for the relative magnifications ($\mu_{21}$ and $\mu_{31}$;
		left column) and relative time delays ($\Delta t_{21}$ and $\Delta t_{32}$;
		right column).
		The indices indicate the signal considered and the reference signal respectively.
		Only posterior samples containing at least two (three) signals are shown
		for $\mu_{21}$ and $\Delta t_{21}$ ($\mu_{31}$ and $\Delta t_{32}$).
		The recovered posteriors are broadly
		consistent across waveform families, favouring a secondary
		signal moderately demagnified relative to the primary and a tertiary
		signal significantly demagnified compared with the primary. The
		\textsc{XO4a} recovery for $\mu_{21}$ exhibits additional support
		for a very faint secondary signal, effectively a single-signal
		configuration, consistent with its small support for
		single-signal configurations.
		The inferred relative time delays
		cluster at $\sim20-40\,\textrm{ms}$, with \textsc{XO4a}
		exhibiting broader support towards larger $\Delta t_{21}$. Overall, the consistency
		of the recovered lensing parameters across waveform models supports the
		robustness of the lensed solution.
	}
	\label{fig:GW231123_lensing_posteriors}
\end{figure}

All waveform models favour the lensed hypothesis, with the support varying in
strength, as quantified by $\log_{10}\mathcal{B}_{\textrm{U}}^{\textrm{L}}$
which ranges from $0.7$ to $4.65$.
The \textsc{NRSur} and \textsc{Phenom} waveforms recover broadly consistent
posterior distributions for the lensing parameters, including the relative
magnifications, time delays, phase shifts, and the preferred number of overlapping signals
(see Extended Data Fig.~\ref{fig:GW231123_lensing_posteriors}), with the exception of
\textsc{XO4a} which favours the two-signal hypothesis by a slim margin over the
three-signal hypothesis favoured by the other waveform models.
This lower support from \textsc{XO4a} is consistent with previous analysis
under the isolated-point mass model, which also found lower support with this waveform
model~\citep{LIGOScientific:2025cwb}.

A previous, much weaker lensing candidate---GW200208\_130117~\citep{abbott2023gwtc}---was examined in detail
in~\citep{Janquart:2023mvf} and was found to most likely be a false candidate
with the isolated point-mass lensing
hypothesis being driven by transient noise in a single detector. To investigate
whether this possibility could be driving the much higher support observed for
GW231123, we perform independent analysis of the Hanford and Livingston data
using the \textsc{XPNR} model.
These analyses reveal consistency in both the support for the lensing hypothesis
and the parameters recovered from each detector, indicating that the support for the lensing hypothesis is coherent between the detectors.

\begin{figure}
	\centering
	\includegraphics[width=\textwidth]{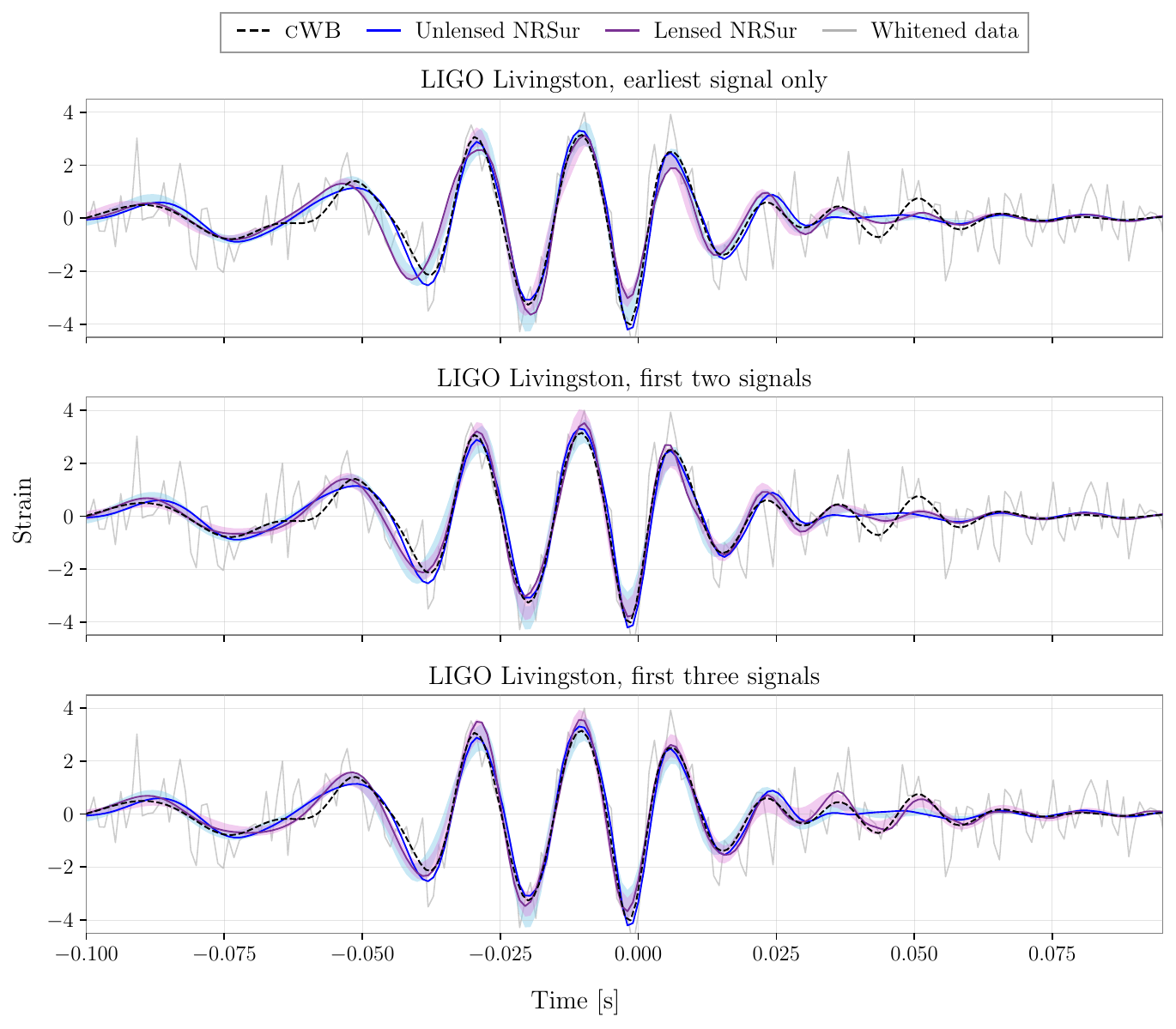}
	\caption{
		\textbf{Extended data}
		Time-domain \textsc{NRSur} reconstruction of GW231123 in the LIGO
		Livingston detector for fixed maximum numbers of lensed signals: one (top), two (middle) and three signals (bottom).
		The one-signal reconstruction largely overlaps with the unlensed
		\textsc{NRSur} reconstruction, while the three-signal solution
		exhibits improved qualitative agreement with the \textsc{cWB}
		reconstruction, particularly in the post-merger signal morphology.
	}
	\label{fig:waveform_k1_k3_images}
\end{figure}

Furthermore, since the flexibility to include additional images may more
easily fit fluctuations in the data, we also explicitly examine cases with a fixed number of signals. These analyses show an increase in the
likelihood as the number of signals is increased from $1$ to $3$ before
saturating (Extended Data Fig.~\ref{fig:GW231123_lensing_posteriors}), indicating that there is diminished support for
additional lensed components beyond a potential third signal.
The corresponding time-domain reconstructions are shown in Extended Data Fig.~\ref{fig:waveform_k1_k3_images}.
Within the phenomenological model, each signal is a time- and phase-shifted
copy of the primary signal with its amplitude scaled by $\sqrt{\mu_{j1}}$, such
that the optimal \gls{snr} each signal would have in isolation is
$\sqrt{\mu_{j1}}$ times that of the primary signal. Evaluating this over the
\textsc{XPNR} posterior samples containing at least three signals, we find
network optimal \glspl{snr} of $18^{+3}_{-3}$, $10^{+4}_{-4}$ and
$3.9^{+2.0}_{-1.7}$ for the primary, secondary and tertiary signals
respectively. As expected for a strongly demagnified signal, the tertiary
signal is too weak to be identified in isolation, but its inclusion is
favoured within the joint model, with $\log_{10}\mathcal{B} \approx 0.8$
between the analyses with three and two fixed signals for each waveform model.
However, the support for lensing does not rely on the tertiary signal: the
two-signal analyses alone yield $\log_{10}\mathcal{B}^{\rm L}_{\rm U} = 2.7$
(\textsc{XPNR}) and $2.9$ (\textsc{NRSur}).

The recovered Morse phases provide an additional consistency check on this
interpretation. Recovering the relative difference between the Morse phases of
each image, we find that the first two images are consistent with having the
same image type, whilst the third is consistent with a different type. This is
an allowed scenario under the multi-image lensing
interpretation~\citep{Schneider:1992bmb}.

It must be noted that the absence of strong evidence for additional
phenomenological components does not exclude physical configurations with more
images. In particular, an additional fourth image could be present but too weak
to be robustly recovered, thus physical four-or-more image configurations may be
represented effectively by a lower-image phenomenological model.

\subsection*{Source properties under the lensing hypothesis}

\begin{figure}
	\centering
	\includegraphics[width=\textwidth]{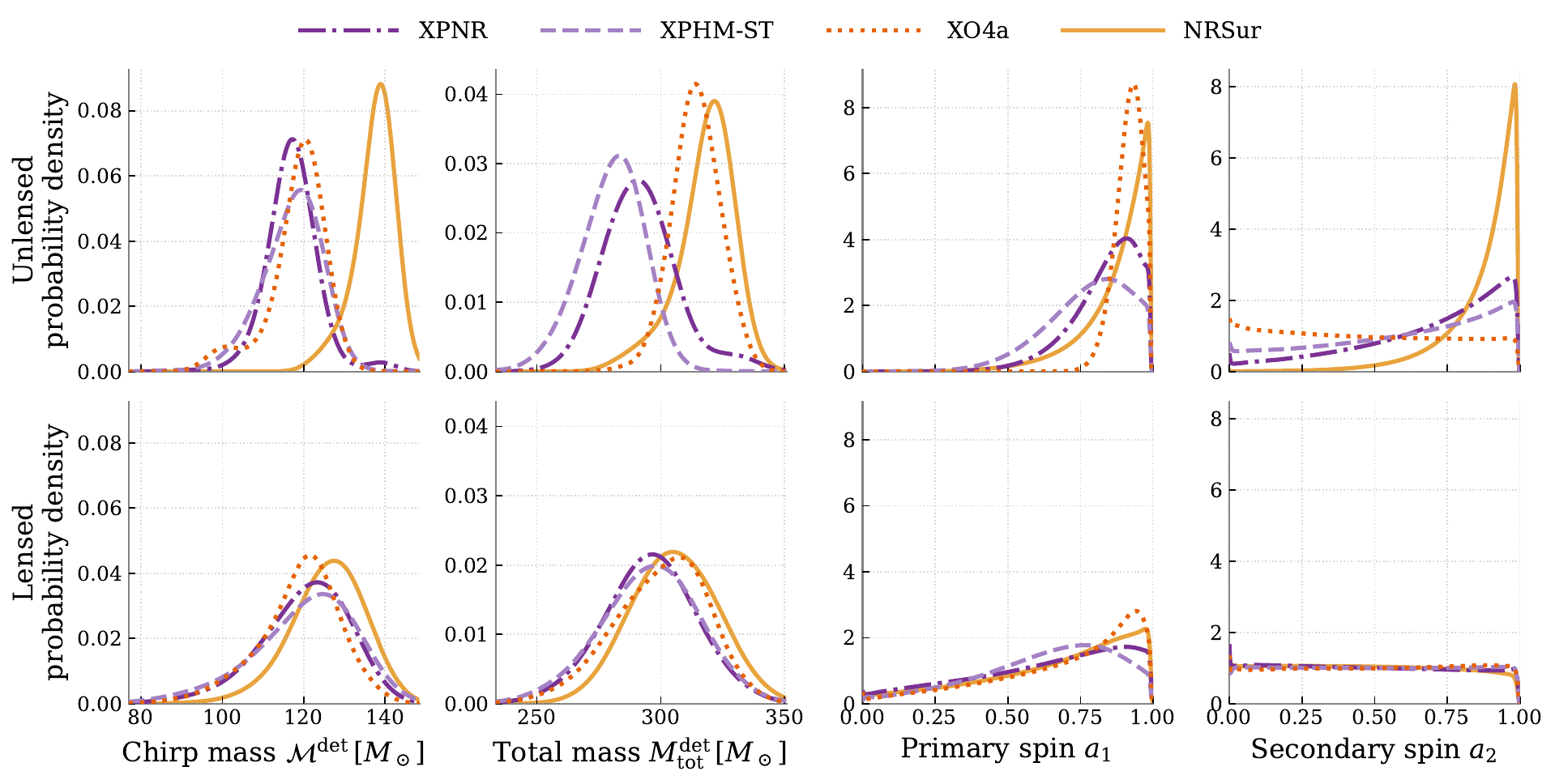}
	\caption{
		Posterior distributions for the detector-frame chirp mass
		($\mathcal{M}^{\rm det}$), detector-frame total mass ($M_{\rm tot}^{\rm det}$),
		and dimensionless spin magnitudes of the primary ($a_1$) and secondary ($a_2$)
		black holes (columns, left to right), recovered under the unlensed (top row)
		and lensed (bottom row) hypotheses with different waveform models.
		Panels in the same column share the same axes.
		Across all waveform models,
		the lensed interpretation yields mutually consistent mass posteriors and
		systematically reduced spin magnitudes relative to the unlensed recovery.
		This indicates GW231123 is more consistent with the bulk of observed \gls{bbh}
		population under the lensed hypothesis.
	}
	\label{fig:masses_and_spins}
\end{figure}

Analyses under the lensing hypothesis systematically shift several inferred
source properties of GW231123 relative to their unlensed
counterparts~\citep{LIGOScientific:2025cwb}. Across all waveform models, the
lensing interpretation yields mutually consistent detector-frame masses, with the
chirp-mass posteriors of the different waveform models converging to
$\mathcal{M}^{\rm det} \approx 120$--$126\,M_\odot$, and reduced
support for extremal spin magnitudes (Fig.~\ref{fig:masses_and_spins}). This
returns the inferred parameters to better-calibrated regions of the parameter
space, which could lead to the improved agreement between different waveform models.
Indeed, under the lensing interpretation, the inferred source properties become substantially more consistent across both the \textsc{NRSur} and \textsc{Phenom} waveform families.
This is also consistent with the interpretation that part of the waveform-model dependence observed under the unlensed hypothesis may arise from missing physical effects---including lensing-induced modulations---being absorbed differently by different waveform approximants.

\begin{figure}
	\centering
	\includegraphics[width=\textwidth]{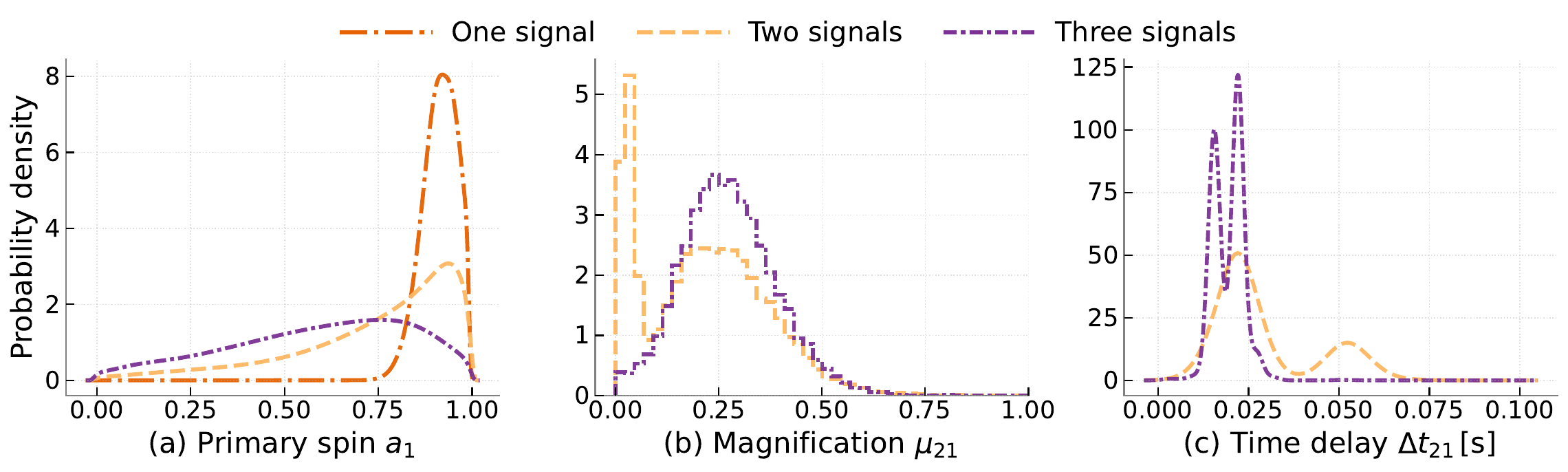}
	\caption{\textbf{Extended data}
		Posterior distributions for (a) the primary spin magnitude $a_1$, (b) the
		relative magnification $\mu_{21}$ and (c) the relative time delay
		$\Delta t_{21}$ obtained with \textsc{XO4a}, with posterior samples separated
		by the number of lensed signals they contain (one, two or three). For a single
		signal, $\mu_{21}$ and $\Delta t_{21}$ do not enter the waveform and are not
		shown. Single-signal samples favour nearly extremal spins, broadly consistent
		with the unlensed inference, suggesting that spin-precession effects may
		partially compensate for the missing additional signals; samples with two or
		three signals show broader support at lower spin magnitudes. The two-signal
		posterior is bimodal. One mode ($\mu_{21} \approx 0.26$,
		$\Delta t_{21} \approx 15$--$28$\,ms) corresponds to the secondary signal of
		the three-signal solution. The other ($\mu_{21} \approx 0.03$,
		$\Delta t_{21} \approx 46$--$61$\,ms) matches its faint tertiary signal
		($\mu_{31} \approx 0.04$, $\Delta t_{31} \approx 57$\,ms), with nearly
		extremal spins as in the single-signal case.
	}
	\label{fig:xo4a}
\end{figure}

Strong support for extremal spin values under the unlensed
hypothesis is also expected for lensed signals analysed without accounting for
lensing effects. Previous work~\citep{Liu:2023emk} identified a
degeneracy between \gls{gw} lensing and spin-induced precession, specifically
showing that unlensed waveform models may partially absorb lensing-induced
amplitude and phase modulations through enhanced spin precession. The reduced
support for extremal spins, together with the increased support for moderate spins
recovered under the lensing hypothesis in this work, is more
consistent with the majority of observed \gls{bbh}
events~\citep{LIGOScientific:2025pvj}.
The differences in treatment of spin dynamics in \textsc{XO4a} relative to the
other waveform models, may thus also contribute to the weaker support for the
lensing hypothesis obtained under this waveform. This is further attested
by the reduction in support for higher spins observed when
allowing for more lensed images (see Extended Data Fig.~\ref{fig:xo4a} for \textsc{XO4a}).

Furthermore, the source-frame masses would be expected to decrease if the signal
is magnified due to lensing. The overall magnification of the signal, i.e. the
amplification of the primary signal relative to an unlensed one, is degenerate
with the luminosity distance and is therefore not measured, neither in an
unlensed analysis nor in our lensed analyses. A signal magnified by lensing
appears closer than it actually is, resulting in an underestimated redshift and
therefore an over-estimation of the source-frame masses.
This would further increase the consistency of GW231123 with the population of
observed \gls{gw} events under the lensed hypothesis (see e.g., Ref~\citep{Goyal:2025eqo}).

\subsection*{Significance against unlensed hypothesis}

\begin{figure}
	\centering
	\includegraphics[width=\textwidth]{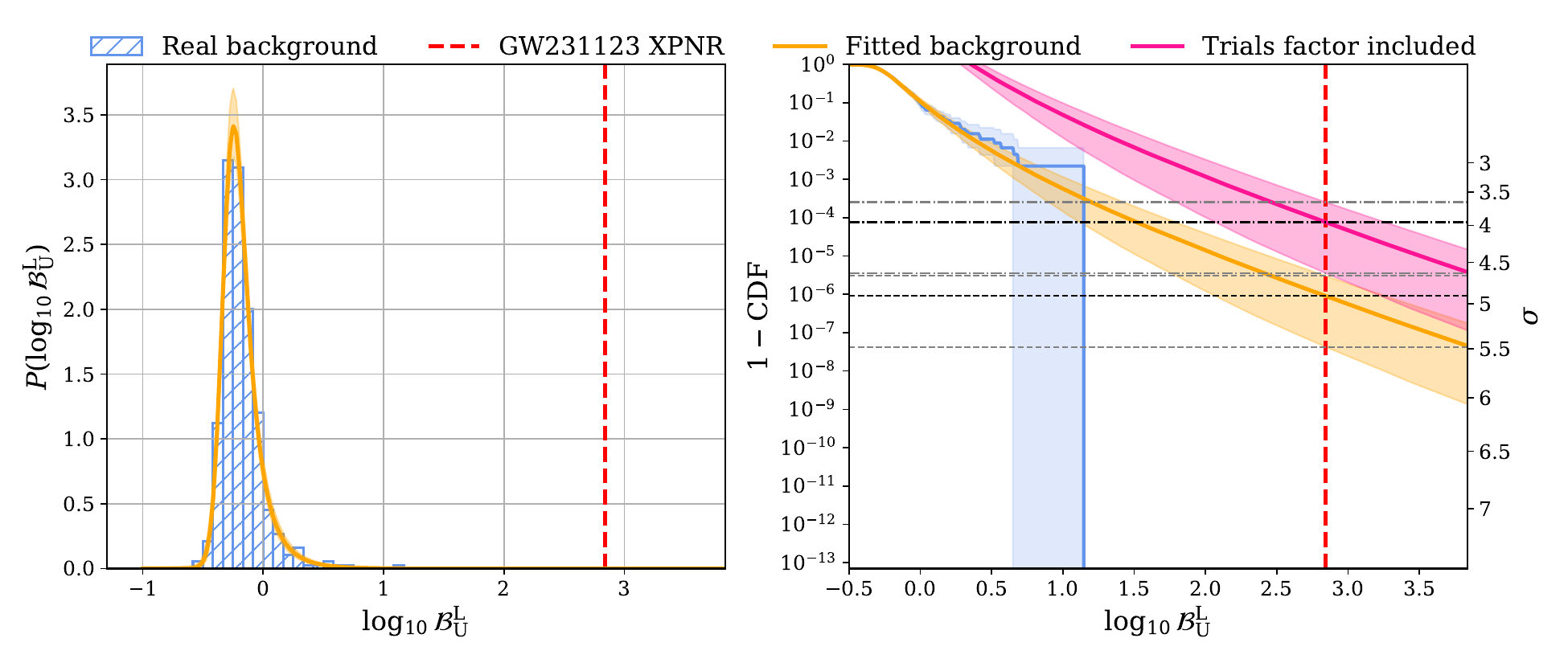}
	\caption{
		\textit{Left:} Histogram distribution of $\log_{10}
			\mathcal{B}^{\textrm{L}}_{\textrm{U}}$ values obtained from the events in
		the background of 450 \textsc{XPNR} injections. Overlaid is the fit
		obtained using the method described in~\citep{Barsode:2026bcs}. The red
		vertical line indicates the value obtained for GW231123 using
		\textsc{XPNR}. \textit{Right:} The complementary cumulative density
		function from the background with the fit values overlaid in orange. The
		pink curve corrects these values for the trials factor considering all
		new \gls{bbh} events in GWTC-4.0 reported from the first part of the fourth
		observing run. Highlighted in the horizontal lines are the
		estimated sigma values for GW231123 at the median of the fit and the 90\%
		CI boundaries.
	}
	\label{fig:background_distribution}
\end{figure}

A background study of $450$ unlensed injections into real detector noise from
the fourth observing run~\citep{LIGOScientific:2025snk} yields an upper bound
on the  \gls{fap} of $0.002\bar{2}$ for the observed Bayes factor comparing the
lensed and unlensed hypotheses for the \textsc{XPNR} analysis. Whilst the
computational expense of running a background consisting of the many more
events necessary to fully calculate the significance of the observed Bayes
factor is too high to reasonably perform without machine-learning methods such
as performed in~\citep{Chan:2025kyu}, the significance can be obtained using
the fitting approach developed in~\citep{Barsode:2026bcs} and shown to match
results obtained with more complete inference schemes. This yields a
significance, under this waveform model, of $3.96^{+0.68}_{-0.30}\sigma$
including a trials factor of 84 to account for the other observed \gls{bbh}
signals added in GWTC-4.0~\citep{LIGOScientific:2025slb}
($4.91^{+0.57}_{-0.24}\sigma$ without), corresponding to the most significant
interpretation of this signal so far. The background distribution, its fit, and
the resulting significance for GW231123 are shown in
Fig.~\ref{fig:background_distribution}. This is also the highest significance
for lensing of a \gls{gw} signal observed to date.  We note that LVK analyses
of GW231123~\citep{LIGOScientific:2025rsn} and the GWTC-4
population~\citep{LIGOScientific:2025pvj} do not identify GW231123 as an
astrophysical population outlier.  This motivates the use of an astrophysical
population for the unlensed background, although high-mass events remain
under-represented in the finite background sample. The waveform model used,
\textsc{XPNR}, as the most recent waveform in the \textsc{Phenom} family and
given its improved calibration to numerical-relativity
simulations~\citep{Hamilton:2025xru}. The significance would vary for other
waveforms in-line with their differing Bayes factors. Waveform systematics also
complicate the interpretation by driving apparent false support for lensing as
noted in~\cite{Hegde:2026anm}---though this support does not typically last
across multiple waveform and lensing models as is the case for GW231123.
However, as this cannot be definitively ruled out, the apparent high
significance is insufficient to make a claim of the detection of lensing from
GW231123.

\section*{Discussion}\label{sec:discussion}
Several alternative explanations have been proposed for the anomalous features
of GW231123, including contamination by glitches~\citep{Ray:2025rtt,
	chatterjee2025machine}, coincident overlapping \gls{bbh}
signals~\citep{Hu:2025lhv}, and orbital eccentricity---which was found to be
disfavoured~\citep{Jan:2025zcm, chandra2026eccentricity}. Many of these share
morphological features with, or are degenerate with, gravitational
lensing~\citep{Romero-Shaw:2022fbf, Divyajyoti:2023rht, Mishra:2025dpa}; in
particular, the apparent deviations reported in the tests of general relativity
on this event~\citep{LIGOScientific:2026qni} can be induced by unmodelled
wave-optics lensing, which therefore offers a natural explanation for these
features~\citep{Mishra:2023vzo, Wright:2024mco, Liu:2024xxn}.

Other lensing analyses of GW231123 have found varying levels of support for the
hypothesis~\citep{Goyal:2025eqo, Shan:2025dcd,
	Chan:2025kyu,Chakraborty:2025pxt}. The significance we have found in this work
is compatible with those estimates. We find that this statistical significance
persists when evaluated against noise which is plausibly non-Gaussian coming
from the detectors directly rather than via simulation. The phenomenological
model probes a much wider array of potential lensing configurations than
analyses based on specific lens models, such as those in
Ref.~\cite{LIGOScientific:2025cwb} or Ref.~\cite{Goyal:2025eqo}. Our results
favour at least a three-signal configuration which would directly discard the
isolated point-mass hypothesis examined in both works. This would instead
favour more complex lens geometries, such as asymmetric or environmentally
perturbed lenses, which were partially examined in Ref.~\cite{Goyal:2025eqo}
which marginal differences in support as compared to isolated models.

Specific lens models would provide further insight into the population of
compact objects capable of producing lensing signatures in \gls{gw}
observatories and would enable a computation of the astrophysical prior odds.
The construction of such odds was briefly discussed in Section 6.4 of
Ref.~\citep{LIGOScientific:2025cwb} and noted as an open question given the
wide array of potential sources that would need to be considered to compute a
robust value. Some investigations with specific potential sources is considered
in~\cite{Vujeva:2026xfh} and~\cite{Cheung:2026pky} which both disfavour the
isolated lens scenario yielding a lensed GW231123.  However, the inference of
lensing parameters in the case of GW231123 is complicated by the absence of a
strongly lensed counterpart signal~\citep{LIGOScientific:2025cwb} which would
allow any macropotential magnification in embedded lensing scenarios to be
disentangled, enabling a full comparison between isolated and embedded models.

The collective evidence from direct analysis of the data in this work
consistently favours a lensing interpretation of GW231123 over the unlensed
merger scenario, with support persistent across multiple waveform models,
between the two observing detectors, and in the presence of actual detector
noise segments. This indicates that the observed structure is unlikely to arise
purely from the presence of unmodelled noise. However, there do remain
modelling uncertainties and limitations on plausible astrophysical sources.
GW231123 therefore provides the strongest evidence to-date for \gls{gw} lensing
in the \gls{lvk} observational era with an estimated significance of
$3.96^{+0.68}_{-0.30}$ based on our semi-analytical fit to an astrophysical
background distribution, however, this support is insufficient for a claim of
lensing given those remaining uncertainties. More broadly, however, these
results indicate that \gls{gw} observations are becoming capable of probing
propagation effects and compact structures beyond the isolated binary-merger
paradigm, opening new avenues for studying compact objects and small-scale
structure through \gls{gw} astronomy.


\begin{thebibliography}{10}
	\expandafter\ifx\csname url\endcsname\relax
		\def\url#1{\burl{#1}}\fi
	\expandafter\ifx\csname urlprefix\endcsname\relax\def\urlprefix{URL }\fi
	\providecommand{\bibinfo}[2]{#2}
	\providecommand{\eprint}[2][]{\url{#2}}
	\providecommand{\doi}[1]{\url{https://doi.org/#1}}
	\bibcommenthead

	\bibitem{LIGOScientific:2025rsn}
	\bibinfo{author}{Abac, A.~G.} \emph{et~al.}
	\newblock \bibinfo{title}{{GW231123: A Binary Black Hole Merger with Total Mass
				190--265 M$_{\odot}$}}.
	\newblock \emph{\bibinfo{journal}{Astrophys. J. Lett.}}
	\textbf{\bibinfo{volume}{993}}, \bibinfo{pages}{L25} (\bibinfo{year}{2025}).

	\bibitem{LIGOScientific:2014pky}
	\bibinfo{author}{Aasi, J.} \emph{et~al.}
	\newblock \bibinfo{title}{{Advanced LIGO}}.
	\newblock \emph{\bibinfo{journal}{Class. Quant. Grav.}}
	\textbf{\bibinfo{volume}{32}}, \bibinfo{pages}{074001}
	(\bibinfo{year}{2015}).

	\bibitem{VIRGO:2014yos}
	\bibinfo{author}{Acernese, F.} \emph{et~al.}
	\newblock \bibinfo{title}{{Advanced Virgo: a second-generation interferometric
				gravitational wave detector}}.
	\newblock \emph{\bibinfo{journal}{Class. Quant. Grav.}}
	\textbf{\bibinfo{volume}{32}}, \bibinfo{pages}{024001}
	(\bibinfo{year}{2015}).

	\bibitem{akutsu2021overview}
	\bibinfo{author}{Akutsu, T.} \emph{et~al.}
	\newblock \bibinfo{title}{Overview of KAGRA: Detector design and construction
		history}.
	\newblock \emph{\bibinfo{journal}{Progress of Theoretical and Experimental
			Physics}} \textbf{\bibinfo{volume}{2021}}, \bibinfo{pages}{05A101}
	(\bibinfo{year}{2021}).

	\bibitem{LIGOScientific:2025pvj}
	\bibinfo{author}{Abac, A.~G.} \emph{et~al.}
	\newblock \bibinfo{title}{{GWTC-4.0: Population Properties of Merging Compact
				Binaries}}  (\bibinfo{year}{2025}).

	\bibitem{Bini:2026kwz}
	\bibinfo{author}{Bini, S.}, \bibinfo{author}{Kr{\'o}l, K.},
	\bibinfo{author}{Chatziioannou, K.} \& \bibinfo{author}{Isi, M.}
	\newblock \bibinfo{title}{{The impact of waveform systematics and Gaussian
				noise on the interpretation of GW231123}}  (\bibinfo{year}{2026}).

	\bibitem{Ray:2025rtt}
	\bibinfo{author}{Ray, A.}, \bibinfo{author}{Banagiri, S.},
	\bibinfo{author}{Thrane, E.} \& \bibinfo{author}{Lasky, P.~D.}
	\newblock \bibinfo{title}{{GW231123: extreme spins or microglitches?}}
	(\bibinfo{year}{2025}).

	\bibitem{Cuceu:2025fzi}
	\bibinfo{author}{Cuceu, I.}, \bibinfo{author}{Bizouard, M.~A.},
	\bibinfo{author}{Christensen, N.} \& \bibinfo{author}{Sakellariadou, M.}
	\newblock \bibinfo{title}{{GW231123: Binary Black Hole Merger or Cosmic
				String?}}  (\bibinfo{year}{2025}).

	\bibitem{Hu:2025lhv}
	\bibinfo{author}{Hu, Q.} \emph{et~al.}
	\newblock \bibinfo{title}{{GW231123: Overlapping Gravitational Wave Signals?}}
	(\bibinfo{year}{2025}).

	\bibitem{LIGOScientific:2023bwz}
	\bibinfo{author}{Abbott, R.} \emph{et~al.}
	\newblock \bibinfo{title}{{Search for Gravitational-lensing Signatures in the
				Full Third Observing Run of the LIGO{\textendash}Virgo Network}}.
	\newblock \emph{\bibinfo{journal}{Astrophys. J.}}
	\textbf{\bibinfo{volume}{970}}, \bibinfo{pages}{191} (\bibinfo{year}{2024}).

	\bibitem{LIGOScientific:2025cwb}
	\bibinfo{author}{Abac, A.~G.} \emph{et~al.}
	\newblock \bibinfo{title}{{GWTC-4.0: Searches for Gravitational-Wave Lensing
				Signatures}}  (\bibinfo{year}{2025}).

	\bibitem{LIGOScientific:2025slb}
	\bibinfo{author}{Abac, A.~G.} \emph{et~al.}
	\newblock \bibinfo{title}{{GWTC-4.0: Updating the Gravitational-Wave Transient
				Catalog with Observations from the First Part of the Fourth LIGO-Virgo-KAGRA
				Observing Run}}  (\bibinfo{year}{2025}).

	\bibitem{Goyal:2025eqo}
	\bibinfo{author}{Goyal, S.}, \bibinfo{author}{Villarrubia-Rojo, H.} \&
	\bibinfo{author}{Zumalacarregui, M.}
	\newblock \bibinfo{title}{{Across the Universe: GW231123 as a magnified and
				diffracted black hole merger}}  (\bibinfo{year}{2025}).

	\bibitem{Shan:2025dcd}
	\bibinfo{author}{Shan, X.}, \bibinfo{author}{Yang, H.} \& \bibinfo{author}{Mao,
		S.}
	\newblock \bibinfo{title}{{GW231123: A Case for Binary Microlensing in a Strong
				Lensing Field}}  (\bibinfo{year}{2025}).

	\bibitem{Chakraborty:2025pxt}
	\bibinfo{author}{Chakraborty, A.} \& \bibinfo{author}{Mukherjee, S.}
	\newblock \bibinfo{title}{{The First Model-Independent Upper Bound on
				Micro-lensing Signature of the Highest Mass Binary Black Hole Event
				GW231123}}  (\bibinfo{year}{2025}).

	\bibitem{Chan:2025kyu}
	\bibinfo{author}{Chan, J. C.~L.} \emph{et~al.}
	\newblock \bibinfo{title}{{Discovering gravitational waveform distortions from
				lensing: a deep dive into GW231123}}  (\bibinfo{year}{2025}).

	\bibitem{Paynter_2021}
	\bibinfo{author}{Paynter, J.}, \bibinfo{author}{Webster, R.} \&
	\bibinfo{author}{Thrane, E.}
	\newblock \bibinfo{title}{Evidence for an intermediate-mass black hole from a
		gravitationally lensed gamma-ray burst}.
	\newblock \emph{\bibinfo{journal}{Nature Astronomy}}
	\textbf{\bibinfo{volume}{5}}, \bibinfo{pages}{560–568}
	(\bibinfo{year}{2021}).
	\newblock \urlprefix\url{http://dx.doi.org/10.1038/s41550-021-01307-1}.

	\bibitem{levan2025gravitational}
	\bibinfo{author}{Levan, A.~J.} \emph{et~al.}
	\newblock \bibinfo{title}{Gravitational lensing in gamma-ray bursts}.
	\newblock \emph{\bibinfo{journal}{Philosophical Transactions of the Royal
			Society A: Mathematical, Physical and Engineering Sciences}}
	\textbf{\bibinfo{volume}{383}} (\bibinfo{year}{2025}).

	\bibitem{vujeva2025dark}
	\bibinfo{author}{Vujeva, L.}, \bibinfo{author}{Ezquiaga, J.~M.},
	\bibinfo{author}{Gilman, D.}, \bibinfo{author}{Goyal, S.} \&
	\bibinfo{author}{Zumalac{\'a}rregui, M.}
	\newblock \bibinfo{title}{Dark matter subhalos and higher order catastrophes in
		gravitational wave lensing}.
	\newblock \emph{\bibinfo{journal}{arXiv preprint arXiv:2510.14953}}
	(\bibinfo{year}{2025}).

	\bibitem{Liu:2023ikc}
	\bibinfo{author}{Liu, A.} \emph{et~al.}
	\newblock \bibinfo{title}{{Exploring the hidden Universe: a novel
				phenomenological approach for recovering arbitrary gravitational-wave
				millilensing configurations}}.
	\newblock \emph{\bibinfo{journal}{Mon. Not. Roy. Astron. Soc.}}
	\textbf{\bibinfo{volume}{525}}, \bibinfo{pages}{4149--4160}
	(\bibinfo{year}{2023}).

	\bibitem{Varma:2019csw}
	\bibinfo{author}{Varma, V.} \emph{et~al.}
	\newblock \bibinfo{title}{{Surrogate models for precessing binary black hole
				simulations with unequal masses}}.
	\newblock \emph{\bibinfo{journal}{Phys. Rev. Research.}}
	\textbf{\bibinfo{volume}{1}}, \bibinfo{pages}{033015} (\bibinfo{year}{2019}).

	\bibitem{Thompson:2023ase}
	\bibinfo{author}{Thompson, J.~E.} \emph{et~al.}
	\newblock \bibinfo{title}{{PhenomXO4a: a phenomenological gravitational-wave
				model for precessing black-hole binaries with higher multipoles and
				asymmetries}}.
	\newblock \emph{\bibinfo{journal}{Phys. Rev. D}}
	\textbf{\bibinfo{volume}{109}}, \bibinfo{pages}{063012}
	(\bibinfo{year}{2024}).

	\bibitem{Pratten:2020ceb}
	\bibinfo{author}{Pratten, G.} \emph{et~al.}
	\newblock \bibinfo{title}{{Computationally efficient models for the dominant
				and subdominant harmonic modes of precessing binary black holes}}.
	\newblock \emph{\bibinfo{journal}{Phys. Rev. D}}
	\textbf{\bibinfo{volume}{103}}, \bibinfo{pages}{104056}
	(\bibinfo{year}{2021}).

	\bibitem{Colleoni:2024knd}
	\bibinfo{author}{Colleoni, M.}, \bibinfo{author}{Vidal, F. A.~R.},
	\bibinfo{author}{Garc{\'\i}a-Quir{\'o}s, C.}, \bibinfo{author}{Ak{\c{c}}ay,
		S.} \& \bibinfo{author}{Bera, S.}
	\newblock \bibinfo{title}{{Fast frequency-domain gravitational waveforms for
				precessing binaries with a new twist}}.
	\newblock \emph{\bibinfo{journal}{Phys. Rev. D}}
	\textbf{\bibinfo{volume}{111}}, \bibinfo{pages}{104019}
	(\bibinfo{year}{2025}).

	\bibitem{Hamilton:2025xru}
	\bibinfo{author}{Hamilton, E.} \emph{et~al.}
	\newblock \bibinfo{title}{{PhenomXPNR: An improved gravitational wave model
				linking precessing inspirals and NR-calibrated merger-ringdown}}
	(\bibinfo{year}{2025}).

	\bibitem{Drago:2020kic}
	\bibinfo{author}{Drago, M.} \emph{et. al}
	\newblock \bibinfo{title}{Coherent WaveBurst, a pipeline for unmodeled
		gravitational-wave data analysis}.
	\newblock \emph{\bibinfo{journal}{SoftwareX}}
	\textbf{\bibinfo{volume}{14}}, \bibinfo{pages}{100678}
	(\bibinfo{year}{2021})

	\bibitem{abbott2023gwtc}
	\bibinfo{author}{Abbott, R.} \emph{et~al.}
	\newblock \bibinfo{title}{GWTC-3: Compact binary coalescences observed by ligo
		and virgo during the second part of the third observing run}.
	\newblock \emph{\bibinfo{journal}{Physical Review X}}
	\textbf{\bibinfo{volume}{13}}, \bibinfo{pages}{041039}
	(\bibinfo{year}{2023}).

	\bibitem{Janquart:2023mvf}
	\bibinfo{author}{Janquart, J.} \emph{et~al.}
	\newblock \bibinfo{title}{{Follow-up analyses to the O3
	LIGO{\textendash}Virgo{\textendash}KAGRA lensing searches}}.
	\newblock \emph{\bibinfo{journal}{Mon. Not. Roy. Astron. Soc.}}
	\textbf{\bibinfo{volume}{526}}, \bibinfo{pages}{3832--3860}
	(\bibinfo{year}{2023}).

	\bibitem{Schneider:1992bmb}
	\bibinfo{author}{Schenider, P.}, \bibinfo{author}{Ehlers, J.}, \&
	\bibinfo{author}{Falco, E.}
	\newblock \bibinfo{title}{Gravitational Lenses}
	\newblock \emph{\bibinfo{publisher}{Springer}}
	\textbf {\bibinfo{isbn}{978-3-540-66506-9}}
	(\bibinfo{year}{1992})

	\bibitem{Liu:2023emk}
	\bibinfo{author}{Liu, A.} \& \bibinfo{author}{Kim, K.}
	\newblock \bibinfo{title}{Can we discern millilensed gravitational-wave signals
		from signals produced by precessing binary black holes with ground-based
		detectors?}
	\newblock \emph{\bibinfo{journal}{Phys. Rev. D}}
	\textbf{\bibinfo{volume}{110}}, \bibinfo{pages}{123008}
	(\bibinfo{year}{2024}).
	\newblock \urlprefix\url{https://link.aps.org/doi/10.1103/PhysRevD.110.123008}.

	\bibitem{LIGOScientific:2025snk}
	\bibinfo{author}{Abac, A.~G.} \emph{et~al.}
	\newblock \bibinfo{title}{{Open Data from LIGO, Virgo, and KAGRA through the
				First Part of the Fourth Observing Run}}  (\bibinfo{year}{2025}).

	\bibitem{Barsode:2026bcs}
	\bibinfo{author}{Barsode, A.}
	\newblock \bibinfo{title}{{BB plot: A Tool for Accurate Model Selection Using
				Bayes factors}}  (\bibinfo{year}{2026}).

	\bibitem{Hegde:2026anm}
	\bibinfo{author}{Hegde, D.} \& \bibinfo{author}{Janquart, J.}
	\newblock \bibinfo{title}{On the interplay between waveform systematics and
		lensing signatures in gravitational-wave signals}
	\newblock \emph{\bibinfo{journal}{arXiv preprint arXiv:2609.03935}}
	(\bibinfo{year}{2026})

	\bibitem{chatterjee2025machine}
	\bibinfo{author}{Chatterjee, C.}, \bibinfo{author}{McGowan, K.},
	\bibinfo{author}{Deshmukh, S.}, \bibinfo{author}{Tyler-Howard, N.} \&
	\bibinfo{author}{Jani, K.}
	\newblock \bibinfo{title}{Machine learning confirms gw231123 is a ``lite''
		intermediate mass black hole merger}.
	\newblock \emph{\bibinfo{journal}{The Astrophysical Journal Letters}}
	\textbf{\bibinfo{volume}{995}}, \bibinfo{pages}{L6} (\bibinfo{year}{2025}).

	\bibitem{Jan:2025zcm}
	\bibinfo{author}{Jan, A.}, \bibinfo{author}{Nicolella, S.},
	\bibinfo{author}{Shoemaker, D.} \& \bibinfo{author}{O'Shaughnessy, R.}
	\newblock \bibinfo{title}{{Measuring Eccentricity and Addressing Waveform
				Systematics in GW231123}}  (\bibinfo{year}{2025}).

	\bibitem{chandra2026eccentricity}
	\bibinfo{author}{Chandra, K.}, \bibinfo{author}{Fernandes, J.},
	\bibinfo{author}{Mittal, A.} \& \bibinfo{author}{Carullo, G.}
	\newblock \bibinfo{title}{Eccentricity in disguise? Insights from GW231123 and
		numerically simulated binary black hole merger signals}.
	\newblock \emph{\bibinfo{journal}{arXiv preprint arXiv:2606.15150}}
	(\bibinfo{year}{2026}).

	\bibitem{Romero-Shaw:2022fbf}
	\bibinfo{author}{Romero-Shaw, I.~M.}, \bibinfo{author}{Gerosa, D.} \&
	\bibinfo{author}{Loutrel, N.}
	\newblock \bibinfo{title}{{Eccentricity or spin precession? Distinguishing
				subdominant effects in gravitational-wave data}}.
	\newblock \emph{\bibinfo{journal}{Mon. Not. Roy. Astron. Soc.}}
	\textbf{\bibinfo{volume}{519}}, \bibinfo{pages}{5352--5357}
	(\bibinfo{year}{2023}).

	\bibitem{Divyajyoti:2023rht}
	\bibinfo{author}{Divyajyoti}, \bibinfo{author}{Kumar, S.},
	\bibinfo{author}{Tibrewal, S.}, \bibinfo{author}{Romero-Shaw, I.~M.} \&
	\bibinfo{author}{Mishra, C.~K.}
	\newblock \bibinfo{title}{{Blind spots and biases: The dangers of ignoring
				eccentricity in gravitational-wave signals from binary black holes}}.
	\newblock \emph{\bibinfo{journal}{Phys. Rev. D}}
	\textbf{\bibinfo{volume}{109}}, \bibinfo{pages}{043037}
	(\bibinfo{year}{2024}).

	\bibitem{Mishra:2025dpa}
	\bibinfo{author}{Mishra, A.} \& \bibinfo{author}{Ganguly, A.}
	\newblock \bibinfo{title}{{Can Eccentric Binary Black Hole Signals Mimic
				Gravitational-Wave Microlensing?}}  (\bibinfo{year}{2025}).

	\bibitem{LIGOScientific:2026qni}
	\bibinfo{author}{Abac, A.~G.} \emph{et~al.}
	\newblock \bibinfo{title}{{GWTC-4.0: Tests of General Relativity. I. Overview
				and General Tests}}  (\bibinfo{year}{2026}).

	\bibitem{Mishra:2023vzo}
	\bibinfo{author}{Mishra, A.}, \bibinfo{author}{Krishnendu, N.~V.} \&
	\bibinfo{author}{Ganguly, A.}
	\newblock \bibinfo{title}{{Unveiling microlensing biases in testing general
				relativity with gravitational waves}}.
	\newblock \emph{\bibinfo{journal}{Phys. Rev. D}}
	\textbf{\bibinfo{volume}{110}}, \bibinfo{pages}{084009}
	(\bibinfo{year}{2024}).

	\bibitem{Wright:2024mco}
	\bibinfo{author}{Wright, M.}, \bibinfo{author}{Janquart, J.} \&
	\bibinfo{author}{Johnson-McDaniel, N.~K.}
	\newblock \bibinfo{title}{{Effect of Deviations from General Relativity on
				Searches for Gravitational-wave Microlensing and Type II Strong Lensing}}.
	\newblock \emph{\bibinfo{journal}{Astrophys. J.}}
	\textbf{\bibinfo{volume}{981}}, \bibinfo{pages}{133} (\bibinfo{year}{2025}).

	\bibitem{Liu:2024xxn}
	\bibinfo{author}{Liu, A.}, \bibinfo{author}{Chandramouli, R.~S.},
	\bibinfo{author}{Hannuksela, O.~A.}, \bibinfo{author}{Yunes, N.} \&
	\bibinfo{author}{Li, T. G.~F.}
	\newblock \bibinfo{title}{{Millilensing induced systematic biases in
				parametrized tests of general relativity}}.
	\newblock \emph{\bibinfo{journal}{Phys. Rev. D}}
	\textbf{\bibinfo{volume}{112}}, \bibinfo{pages}{064043}
	(\bibinfo{year}{2025}).

	\bibitem{Vujeva:2026xfh}
	\bibinfo{author}{Vujeva, L.}, \bibinfo{author}{Ezquiaga, J. M.},
	\bibinfo{author}{Lo, R. K. L.}, \& \bibinfo{Zwick, L.}
	\newblock \bibinfo{title}{Probability of gravitational-wave lensing by
		intermediate-mass black holes and globular clusters}
	\newblock \emph{\bibinfo{journal}{arXiv preprint arXiv:2608.07667}}
	(\bibinfo{year}{2026})

	\bibitem{Cheung:2026pky}
	\bibinfo{author}{Cheung, M. H. Y.}, \bibinfo{author}{Wadekar, D.},
	\bibinfo{author}{Zaldarriga, M.}, \bibinfo{author}{Venumadhav, T.},
	\bibinfo{author}{Roulet, J.}, \& \bibinfo{author}{Mehta, A. J}
	\newblock \bibinfo{title}{The diffraction-lensing interpretation of GW231123
		with astrophysical priors}
	\newblock \emph{\bibinfo{journal}{arXiv preprint arXiv:2607.21834}}
	(\bibinfo{year}{2026})
\end{thebibliography}

\begin{thebibliography}{10}
	\expandafter\ifx\csname url\endcsname\relax
		\def\url#1{\burl{#1}}\fi
	\expandafter\ifx\csname urlprefix\endcsname\relax\def\urlprefix{URL }\fi
	\providecommand{\bibinfo}[2]{#2}
	\providecommand{\eprint}[2][]{\url{#2}}
	\providecommand{\doi}[1]{\url{https://doi.org/#1}}
	\makeatletter\setcounter{NAT@ctr}{45}\makeatother


	\bibitem{Takahashi:2003ix}
	\bibinfo{author}{Takahashi, R.} \& \bibinfo{author}{Nakamura, T.}
	\newblock \bibinfo{title}{{Wave effects in gravitational lensing of
				gravitational waves from chirping binaries}}.
	\newblock \emph{\bibinfo{journal}{Astrophys. J.}}
	\textbf{\bibinfo{volume}{595}}, \bibinfo{pages}{1039--1051}
	(\bibinfo{year}{2003}).

	\bibitem{Wright:2021cbn}
	\bibinfo{author}{Wright, M.} \& \bibinfo{author}{Hendry, M.}
	\newblock \bibinfo{title}{{Gravelamps: Gravitational Wave Lensing Mass Profile
				Model Selection}}  (\bibinfo{year}{2021}).

	\bibitem{Ashton:2018jfp}
	\bibinfo{author}{Ashton, G.} \emph{et~al.}
	\newblock \bibinfo{title}{{BILBY: A user-friendly Bayesian inference library
				for gravitational-wave astronomy}}.
	\newblock \emph{\bibinfo{journal}{Astrophys. J. Suppl.}}
	\textbf{\bibinfo{volume}{241}}, \bibinfo{pages}{27} (\bibinfo{year}{2019}).

	\bibitem{Romero-Shaw:2020owr}
	\bibinfo{author}{Romero-Shaw, I.~M.} \emph{et~al.}
	\newblock \bibinfo{title}{{Bayesian inference for compact binary coalescences
				with bilby: validation and application to the first LIGO{\textendash}Virgo
				gravitational-wave transient catalogue}}.
	\newblock \emph{\bibinfo{journal}{Mon. Not. Roy. Astron. Soc.}}
	\textbf{\bibinfo{volume}{499}}, \bibinfo{pages}{3295--3319}
	(\bibinfo{year}{2020}).

	\bibitem{Speagle:2019ivv}
	\bibinfo{author}{Speagle, J.~S.}
	\newblock \bibinfo{title}{{dynesty: a dynamic nested sampling package for
				estimating Bayesian posteriors and evidences}}.
	\newblock \emph{\bibinfo{journal}{Mon. Not. Roy. Astron. Soc.}}
	\textbf{\bibinfo{volume}{493}}, \bibinfo{pages}{3132--3158}
	(\bibinfo{year}{2020}).

	\bibitem{Robinet:2020lbf}
	\bibinfo{author}{Robinet, F.} \emph{et~al.}
	\newblock \bibinfo{title}{{Omicron: a tool to characterize transient noise in
				gravitational-wave detectors}}.
	\newblock \emph{\bibinfo{journal}{SoftwareX}} \textbf{\bibinfo{volume}{12}},
	\bibinfo{pages}{100620} (\bibinfo{year}{2020}).

	\bibitem{Cornish:2011ys}
	\bibinfo{author}{Cornish, N.}, \bibinfo{author}{Sampson, L.},
	\bibinfo{author}{Yunes, N.} \& \bibinfo{author}{Pretorius, F.}
	\newblock \bibinfo{title}{{Gravitational Wave Tests of General Relativity with
				the Parameterized Post-Einsteinian Framework}}.
	\newblock \emph{\bibinfo{journal}{Phys. Rev. D}} \textbf{\bibinfo{volume}{84}},
	\bibinfo{pages}{062003} (\bibinfo{year}{2011}).

	\bibitem{Vallisneri:2012qq}
	\bibinfo{author}{Vallisneri, M.}
	\newblock \bibinfo{title}{{Testing general relativity with gravitational waves:
				a reality check}}.
	\newblock \emph{\bibinfo{journal}{Phys. Rev. D}} \textbf{\bibinfo{volume}{86}},
	\bibinfo{pages}{082001} (\bibinfo{year}{2012}).

	\bibitem{DelPozzo:2014cla}
	\bibinfo{author}{Del~Pozzo, W.}, \bibinfo{author}{Grover, K.},
	\bibinfo{author}{Mandel, I.} \& \bibinfo{author}{Vecchio, A.}
	\newblock \bibinfo{title}{{Testing general relativity with compact coalescing
				binaries: comparing exact and predictive methods to compute the Bayes
				factor}}.
	\newblock \emph{\bibinfo{journal}{Class. Quant. Grav.}}
	\textbf{\bibinfo{volume}{31}}, \bibinfo{pages}{205006}
	(\bibinfo{year}{2014}).

	\bibitem{MacKay:2003}
	\bibinfo{author}{MacKay, D.~J.~C.}
	\newblock \emph{\bibinfo{title}{Information Theory, Inference, and Learning
			Algorithms}} (\bibinfo{publisher}{Cambridge University Press},
	\bibinfo{year}{2003}).

	\bibitem{Pickands1975Statistical}
	\bibinfo{author}{Pickands, J.}
	\newblock \bibinfo{title}{Statistical inference using extreme order
		statistics}.
	\newblock \emph{\bibinfo{journal}{Annals of Statistics}}
	\textbf{\bibinfo{volume}{3}}, \bibinfo{pages}{119--131}
	(\bibinfo{year}{1975}).
	\newblock \urlprefix\url{https://api.semanticscholar.org/CorpusID:121269906}.

	\bibitem{Balkema}
	\bibinfo{author}{Balkema, A.~A.} \& \bibinfo{author}{de~Haan, L.}
	\newblock \bibinfo{title}{{Residual Life Time at Great Age}}.
	\newblock \emph{\bibinfo{journal}{The Annals of Probability}}
	\textbf{\bibinfo{volume}{2}}, \bibinfo{pages}{792 -- 804}
	(\bibinfo{year}{1974}).
	\newblock \urlprefix\url{https://doi.org/10.1214/aop/1176996548}.

\end{thebibliography}

\section*{Methods}\label{sec:methods}
\subsection*{A Phenomenological Approach to Lensing}
\label{subsec:phenomenological-lensing}

In this work, we adopt the phenomenological approach to lensing pioneered
by~\cite{Liu:2023ikc}. This approach, under the auspices of the geometric-optics
approximation valid for lenses with masses greater than $\sim 100\, M_\odot$~\citep{Takahashi:2003ix}, models the lensed signal as a coherent
superposition of $k$ lensed images, with each image sharing the same intrinsic
waveform morphology and sky position as the original source signal, but magnified,
delayed in time, and possibly phase-shifted. The strain of the lensed signal,
$\tilde{h}_{L}(f)$, is then related to the strain
of the source signal, $\tilde{h}_{U}(f)$, through a frequency-dependent
amplification factor, $\tilde{F}(f)$, by:
\begin{equation}
	\label{eq:amplification-factor}
	\begin{aligned}
		\tilde{h}_{L}(f) & = \tilde{F}(f)\tilde{h}_{U}(f) \\
		                 & = \sqrt{\mu_{1}} \left[
			\exp (-i\pi n_{1}) + \sum_{j=2}^{k} \sqrt{\mu_{j1}}
			\exp \left(
			2\pi{}if\Delta{}t_{j1} - i\pi{}n_{j}
			\right)
			\right]
		\tilde{h}_{U}(f) \, ,
	\end{aligned}
\end{equation}
where $\mu_{1}$ represents the magnification of the first image which may be
absorbed into its effective luminosity distance, and
$\mu_{\textrm{ji}}$ and $\Delta{}t_{ji}$ represent the relative magnification
and time delay of the $j^{\textrm{th}}$ image relative to the
$i^{\textrm{th}}$. The Morse index, $n_{j}$,
may take one of three discrete values---$0.0, 0.5$, or $1.0$---corresponding to
whether the stationary point of the Fermat potential of the lensing object is a
minimum, saddle point, or maximum respectively.
Consequently, a configuration with a single signal ($k = 1$) differs from the
unlensed model only by the constant Morse phase shift $\exp(-i\pi n_{1})$.

We here consider the scenario that the result of the lensing phenomenon is a
single distorted signal rather than multiple distinctly resolvable signals. As
such, we restrict the time delays to the region shorter than the chirp
timescale of the signal and remaining within the overall signal duration. This
typically is on the order of milliseconds to $\mathcal{O}(0.1)$s. These
durations correspond to lens masses in the range $[10^{2}, 10^{6}]M_{\odot}$
when considering the sensitive frequency band of the current ground-based
\gls{gw} detector network~\citep{Liu:2023ikc}. Such masses encompass lens
objects such as \glsfmtlongpl{imbh}, dark matter substructures within galactic
haloes, and dense stellar clusters. We note that the astrophysical priors on
lensing resulting from such objects, e.g. those on the intrinsic optical depth
of the lens populations and the properties of the local lens environment,
remain highly uncertain though work has been conducted on certain models
in~\cite{Vujeva:2026xfh} and~\cite{Cheung:2026pky}.

When the impacts of lensing from this scale of lens are not accounted for in
\gls{gw} analysis, biases are incurred in the parameter estimation. In
particular, not accounting for lensing can lead to higher observed values of
effective spin~\citep{Liu:2023emk} and lower estimates of luminosity distance
due to the magnification. These biases complicate the interpretation of
individual events and disentangling them requires dedicated analysis accounting
for both the possible lensing biases themselves and the uncertainty in waveform
models, as done in this work.

Using the model detailed above, we then evaluate whether a signal favours the
lensing hypothesis by comparing the evidence obtained using this model and that
obtained when doing standard (unlensed) parameter estimation. Mathematically,
this is done by taking the ratio of evidence, leading to the Bayes factor
$\mathcal{B}^{\rm L}_{\rm U}$, where lensing is favoured if $\mathcal{B}^{\rm
		L}_{\rm U} > 1$. In this work, analyses are performed using the
\textsc{Gravelamps}~\citep{Wright:2021cbn} package, which employs the
\textsc{bilby}~\citep{Ashton:2018jfp, Romero-Shaw:2020owr} Bayesian inference
framework and \textsc{dynesty} nested sampler~\citep{Speagle:2019ivv}. Analyses
are carried out assuming the same priors on the source parameters as those used in
GWTC-4.0~\citep{LIGOScientific:2025slb}. The priors on the lensing parameters
are chosen to broadly cover the relevant parameter space, with a log-uniform
prior on $\mu_{j1} \in [0.01, 100]$, a uniform prior on $\Delta t_{ji} \in
	[0.001, 0.1]\mathrm{s}$ and discrete uniform prior for Morse phase index $\{0, 0.5, 1\}$.
The prior on the number of images is a discrete power
law between $1$ and $6$. Consecutive signals are ordered such that their
arrival times satisfy $t_{j+1} > t_j$ to prevent label-switching during the
sampling process.
The number of images is not fixed in advance but is inferred together with the
other parameters through transdimensional sampling: it is treated as a discrete
parameter, the lensing parameters of all possible images are sampled, and only
those of the first $k$ images contribute to the signal model in a given sample.
A single analysis therefore explores configurations with different numbers of
images, and the posterior on $k$ reflects their relative support; further
details are given in~\cite{Liu:2023ikc}. In addition, we perform analyses with
the number of images fixed to $k = 1, \ldots, 5$.

\subsection*{Significance Calculation}\label{subsec:background-generation}

To establish the statistical significance of the inferred $\mathcal{B}^{\rm
		L}_{\rm U}$ values for GW231123, we carry out a background injection campaign
allowing the computation of a \gls{fap}. We simulate a population of
\glspl{bbh} with masses drawn according to the \textsc{Broken Power Law + 2
	Peaks}, spins according to the \textsc{Gaussian Component Spins}, and redshifts
according to the \textsc{Power Law Redshift} model fits from
GWTC-4.0~\citep{LIGOScientific:2025pvj}.

The \gls{gw} signals are then generated using the \textsc{XPNR}
approximant, a recent \textsc{Phenom} waveform model better calibrated to
numerical relativity simulations than previous models and computationally more
efficient than \textsc{NRSur}~\cite{Hamilton:2025xru}. Whilst \textsc{NRSur} is
generally considered more robust, it is calibrated only in specific regions of
parameter space rather than across the broader population, further motivating
the use of the \textsc{XPNR} model. The generated signals are then injected
into stretches of detector noise from the first part of the fourth observing
run of the \gls{lvk} network during which one/both of the LIGO Hanford and LIGO
Livingston detectors were operational and in which the Omicron characterization
pipeline~\citep{Robinet:2020lbf} did not detect any excess power representing a
\gls{snr} greater than 6. Once injected, the signals were then subject to a
detection criterion of an optimal \gls{snr} in the network of $8$ and a
single-detector criterion of \gls{snr} $>4.5$. Samples were drawn until $450$
detected signals were available, allowing  an extrapolation of the \gls{fap} to
be made using the approach detailed in~\citep{Barsode:2026bcs} and explained
later.

These signals were then analysed under both the lensed and unlensed hypotheses
to obtain a distribution of $\mathcal{B}^{\mathrm{L}}_{\mathrm{U}}$ values. This empirical
distribution can then be used to calculate the \gls{fap}, which quantifies the
expected probability that the given Bayes factor value could result from the
fluctuations in support arising from the causes outlined above based on our
current best understanding of the \gls{bbh} population. It corresponds to the
ratio between simulations with a Bayes factor higher or equal to the one
observed for the real event, and the total number of simulations done.

In our case, since no simulation has a higher Bayes factor, the value found is an upper bound on
the \gls{fap} of $1/450$, similarly to what was done in Ref.~\cite{LIGOScientific:2025cwb}.
Increasing the significance much higher would become computationally intractable with
traditional parameter estimation approaches. Therefore, here, we obtain the value
of the significance by using the approach developed in Ref.~\citep{Barsode:2026bcs}, and
shown to match well results from large backgrounds. Here we give the main steps
used to perform the interpolation.

Starting from the ``fitting factor'' approximation \cite{Cornish:2011ys,Vallisneri:2012qq,DelPozzo:2014cla},
where the fitting factor $FF$ quantifies how close a lensed waveform lies
to the manifold of unlensed waveforms, we marginalize over $FF$ to obtain the
following approximation for the $\log_{10}$ of the Bayes factor:
\begin{equation}
	\log_{10} \mathcal{B}^\mathrm{L}_\mathrm{U} (\rho) \approx f \rho^2 + d\log_{10} \rho- \mathcal{O} + \mathcal{N}\left(0, s_0 + s_1\rho + s_2 \rho^2\right)\,,
\end{equation}
where $d$ is the number of additional parameters in the lensing
hypothesis and $\mathcal{O}$ is Occam's factor. The Occam factor quantifies the reduction of the prior volume
of the additional lensing parameters due to the information in the data, and
therefore penalizes the more complex lensed model~\citep{MacKay:2003,
	Barsode:2026bcs}. Since the measurement uncertainties of the parameters
decrease roughly in proportion to $\rho$, this \gls{snr} dependence is written
explicitly through the $d\log_{10}\rho$ term, leaving $\mathcal{O}$
approximately constant for a given analysis~\citep{Barsode:2026bcs}. The term
$\mathcal{N}(0, \sigma)$ denotes a zero-mean Gaussian random variable with
standard deviation $\sigma = s_0 + s_1\rho + s_2\rho^2$, which describes the
scatter of the Bayes factor between signals of the same \gls{snr} arising from
noise fluctuations and the varying fitting
factor~\citep{Vallisneri:2012qq, Barsode:2026bcs}.
$d$ and $\mathcal{O}$ are considered free
parameters in this setup, just like $f$, $s_0$, $s_1$ and $s_2$.
Next, assuming a power law distribution $\rho^{-\alpha}$ above a certain
threshold $\rho_\mathrm{th}$ for the SNR, we marginalize over $\rho$ to
obtain a foreground distribution $P(\log_{10} \mathcal{B}^\mathrm{L}_\mathrm{U}\,|\,\mathcal{H}_\mathrm{L})$,
where $\mathcal{H}_\mathrm{L}$ denotes the lensing hypothesis.
Finally, we use the ``Bayes factor-Bayes factor'' (BB)
relationship~\citep{Barsode:2026bcs}
\begin{equation}
	P(\mathcal{B}^\mathrm{L}_\mathrm{U}\,|\,\mathcal{H}_\mathrm{U}) = \frac{P(\log_{10} \mathcal{B}^\mathrm{L}_\mathrm{U}\,|\,\mathcal{H}_\mathrm{L})}{\big(\mathcal{B}^\mathrm{L}_\mathrm{U}\big)^2}
\end{equation}
to convert the foreground to a background. The priors for the free parameters match
those given in Table I of Ref.~\cite{Barsode:2026bcs}, with the exceptions of the Occam
factor and dimension difference ($\mathcal{O}$ and $d$) which are extended to
upper limits of $15$ and $16$ respectively. We can fit this against the distribution
of $\mathcal{B}^\mathrm{L}_\mathrm{U}$ obtained from the background events
to obtain posterior distributions on the parameters $f$, $\alpha$, $\rho_\mathrm{th}$,
$d$, $\mathcal{O}$, $s_0$, $s_1$ and $s_2$ using nested sampling. This allows us to extrapolate the background distribution to
higher values of $\mathcal{B}^\mathrm{L}_\mathrm{U}$ and obtain an estimate of the \gls{fap},
as can be seen in Fig.~\ref{fig:background_distribution}.
This approach relies on high \gls{snr}
and Gaussian stationary noise. However, it was shown to match
well background estimates found by doing extremely large number of injections, as shown
in Ref.~\cite{Barsode:2026bcs}, showing it is a reliable method to extrapolate
the background distribution to high values of $\mathcal{B}^\mathrm{L}_\mathrm{U}$.

\begin{figure}
	\centering
	\includegraphics[width=\textwidth]{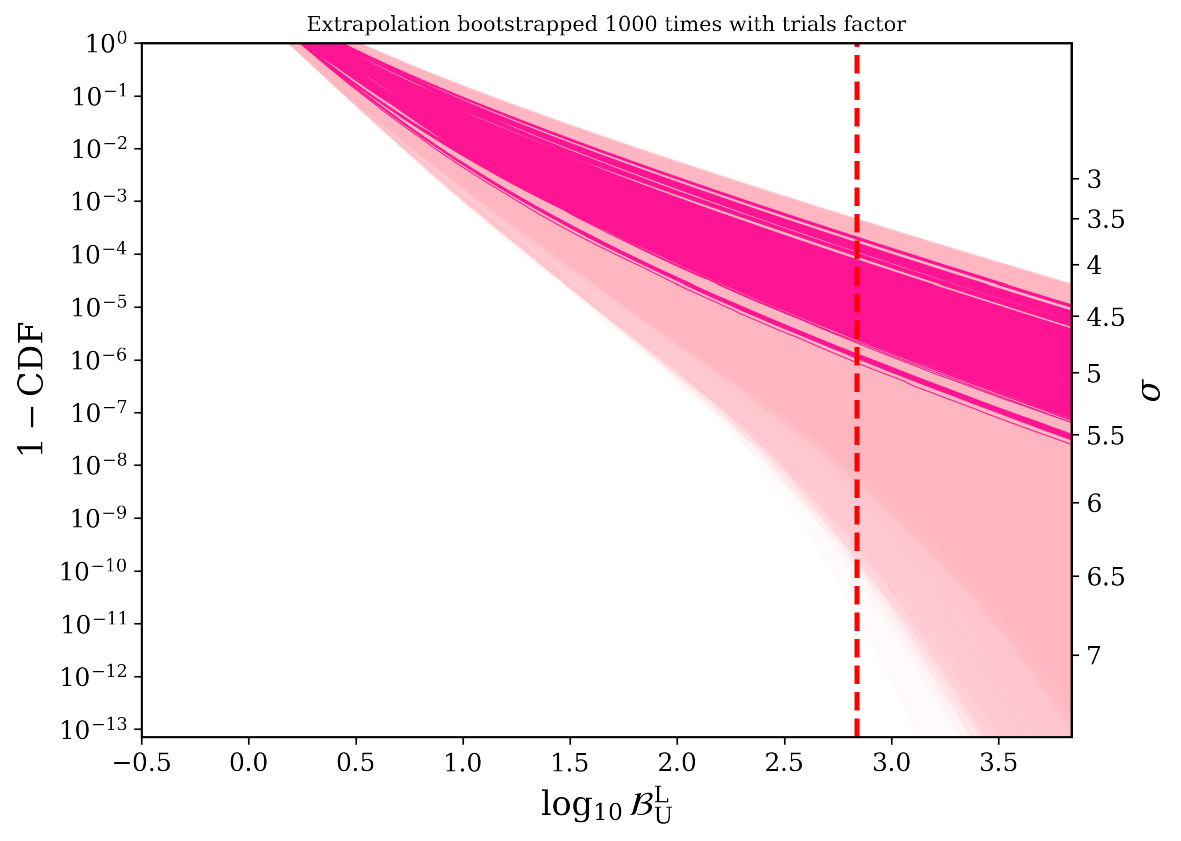}
	\caption{
		\textbf{Extended data}
		Result of bootstrapping the BB-based extrapolation $1000$ times. The deep pink lines correspond to the medians of the posteriors of the fitting parameters. The light pink lines correspond to the $0.05$ and $0.95$ quantiles. The trials factor is already taken into account.
	}
	\label{fig:bootstrapping}
\end{figure}

\begin{figure}
	\centering
	\includegraphics[width=\textwidth]{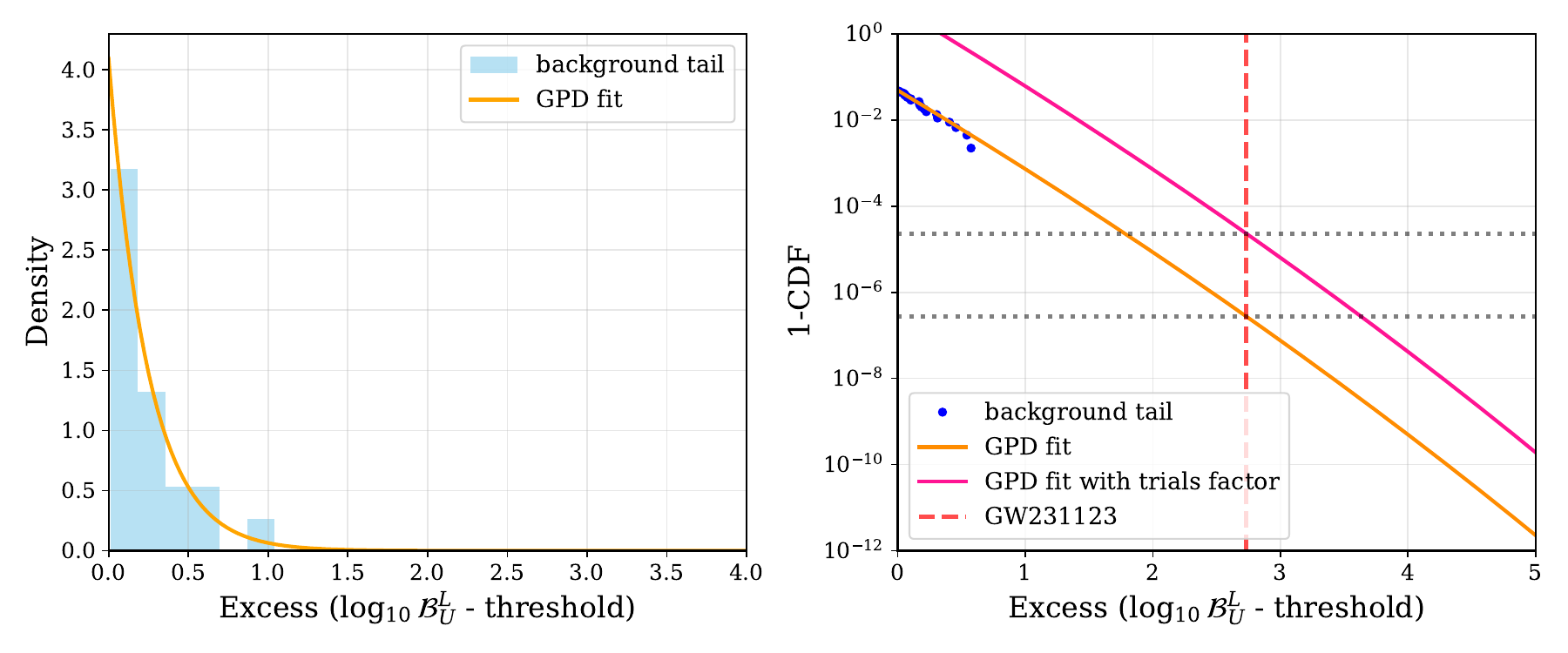}
	\caption{
		\textbf{Extended data}
		Result of fitting the tail of the background with a GPD. The tail is defined by taking the top $5\%$. The threshold is then defined as the lowest $\log_{10}$ Bayes factor of this tail.
	}
	\label{fig:gpdfit}
\end{figure}

To test the robustness of the BB-based extrapolation, we bootstrapped the found
significance by resampling (with replacement) the events in the background
$1000$ times. The result of this is shown in Extended Data
Fig.~\ref{fig:bootstrapping}.
While the upper bound on the significance can be large, the lower bound remains
above approximately $3.51\sigma$ throughout bootstrapping. As a second robustness test, we used the Pickands-Balkema-De Haan theorem~\citep{Pickands1975Statistical, Balkema} as an independent method to extrapolate the background. This theorem, which does not assume Gaussian noise, says that the conditional excess distribution function of a random variable $X$ above a certain threshold $u$, i.e. $F_u(y) = P\left(X-u\leq y\, | \,X > u \right)$, converges to a generalized Pareto distribution (GPD) in the large $u$ limit. In other words, we can approximate the tail of a distribution by a GPD. We take the highest $5\%$
of our background distribution to be the tail, leaving $22$ samples. We then performed a GPD fit using maximum likelihood estimation on this tail and extrapolated to the excess value of GW231123, finding a $\sigma$ value of $4.23$, agreeing remarkably well with the BB-based extrapolation. The result  of the fit is shown in Extended Data Fig.~\ref{fig:gpdfit}

\renewcommand{\refname}{Methods references}

\bmhead{Data availability}

The strain data for GW231123 and the segments of O4a detector noise used
to construct the unlensed background are publicly available from the
Gravitational Wave Open Science Center at
\url{https://gwosc.org}~\citep{LIGOScientific:2025snk}. The calibration
uncertainty envelopes and power spectral densities used in this analysis
form part of the GWTC-4.0 data release~\citep{LIGOScientific:2025slb}.
Posterior samples for all lensed and unlensed analyses of GW231123
presented in this work, together with the injection parameters and the
$\log_{10}\mathcal{B}^{\rm L}_{\rm U}$ values for the 450 background
injections, are available on Zenodo \url{https://doi.org/10.5281/zenodo.21787807}.

\bmhead{Code availability}

The analyses presented here were carried out using
\textsc{Gravelamps}~\citep{Wright:2021cbn}, which employs the
\textsc{bilby} inference framework~\citep{Ashton:2018jfp,
	Romero-Shaw:2020owr} and the \textsc{dynesty} nested
sampler~\citep{Speagle:2019ivv}; all are publicly available. The custom
code used to generate the unlensed background population, perform the
injections, and compute the significance---including the BB-based
extrapolation of Ref.~\citep{Barsode:2026bcs}, the generalized Pareto
tail fit, and the bootstrap---is available on
Zenodo \url{https://doi.org/10.5281/zenodo.21787807}, released under the MIT licence.

\bmhead{Acknowledgements}

The authors are very grateful for discussions with Ankur Barsode and in
particular his support in the computation of the significance values.
The authors thank Anuj Mishra for making available background-generation scripts on which parts of the analysis code were based.
The authors would also like to thank Ajith Parameswaran for his detailed review,
Jose Maria Ezquiaga for feedback on the initial draft of the manuscript,
and the LVK lensing group for useful discussions.
AL would like to thank Nicolas Yunes and Yiqi Xie for discussions and helpful
comments.

\bmhead{Funding}

The authors are grateful for the computing resources provided by the
Consortium des \'{E}quipments (C\'{E}CI) funded by the Fonds de la Recherche
Scientifique de Belgique (F.R.S.-FNRS) under Grant No. 2.4020.11 and by the
Walloon Region and by the LIGO Lab and supported by the National Science
Foundation Grants PHY-0757058 and PHY-0823459. This work used the Dutch
national e-infrastructure with the support of the SURF Cooperative using grant
no. ENIF-16136. This material is based upon work supported by NSF's LIGO
Laboratory which is a major facility fully funded by the National Science
Foundation. HN, MW and CVDB are supported by the research programme of the
Nederlandse Organisatie voor Wetenschappelijk Onderzoek (Netherlands
Organisation for Scientific Research, NWO). AL is partially supported by the
Simons Foundation (via Award No. 896696), the Simons Foundation International
(via Grant No. SFI-MPS-BH-00012593-01), and the NSF (via Grants No.
PHY-2512423). JH is a FRIA grantee of the Fonds de la Recherche Scientifique -
FNRS. JJ acknowledges support from the Fonds de la Recherche Scientifique and
the Royal Observatory of Belgium. ES is supported by grants from the College of Science and Engineering of the University of Glasgow.
LEU is supported by the Hong Kong PhD Fellowship Scheme (HKPFS) from the Hong Kong Research Grants Council (RGC).
AL, LEU, OAH acknowledge support by grants from the Research Grants Council of Hong Kong (Project No. CUHK 14307923), the start-up grant from the Chinese University of Hong Kong, and the Direct Grant for Research from the Research Committee of The Chinese University of Hong Kong.
MH is supported by the UK Science and Technology Facilities Council (Ref. ST/L000946/1).
TGFL is supported by the Research Foundation – Flanders (FWO) under grant number G086722N, I002123N and I000725N.
ICFW is partially supported by the Research Foundation - Flanders (FWO) through Grant No. I002123N.

\section*{Declarations}\label{sec:declarations}
\textbf{Author contributions}

A.L. and M.W. co-led the project; conceived and coordinated the study; co-led
discussions; contributed to the data analysis and development of the Bayesian
framework; performed the assessment of results and figure production; and co-led
the writing and final editing of the manuscript.

J.H. prepared the background frames and contributed to the data analysis of the background, performed the assessment of results and figure production of the background, and contributed to the writing and editing of the manuscript.

H.N. contributed to the data analysis of the background.

E.S. primarily contributed to the methodology writing; participated in
discussions; performed the assessment and validation of the results; and
contributed to the overall review and editing of the manuscript.

I.C.F.W. contributed to the writing of the Results section, participated in discussions and interpretation of the results, and contributed to the review of the results and manuscript.

D.H. contributed to writing up the results and participated in discussions.

LEU contributed to writing up the Discussion section and participated in discussions.

J.J. provided scientific supervision and support throughout the project, contributed to discussion and interpretation of the results, and reviewed and revised the manuscript.

CVDB, MH, TGFL, and OAH supervised the project and contributed to the interpretation of the results and overall scientific guidance.

All authors contributed to editing the manuscript.

\bmhead{Competing interests}

The authors declare no competing interests.


\end{document}